\documentclass{bmvc2k}
\usepackage{graphicx}
\usepackage{amsmath}
\usepackage{amssymb}
\usepackage{booktabs}
\usepackage{multirow}
\usepackage{arydshln}
\usepackage{comment}

\newcommand{\subheader}[1]{\textbf{#1}}

\title{Open Set Synthetic Image Source Attribution}

\addauthor{Shengbang Fang}{sf683@drexel.edu}{1}
\addauthor{Tai D. Nguyen}{tdn47@drexel.edu}{1}
\addauthor{Matthew C. Stamm}{mcs382@drexel.edu}{2}

\addinstitution{
 ECE Department\\
 Drexel University\\
 Philadelphia, PA, USA
}

\runninghead{S.Fang et al.}{Open Set Synthetic Image Source Attribution}

\begin{document}

\maketitle

\begin{abstract}
AI-generated images have become increasingly realistic and have garnered significant public attention.
While synthetic images are intriguing due to their realism, they also pose an important misinformation threat.
To address this new  threat, 
researchers have developed multiple algorithms to detect synthetic images and identify their source generators.
However, most existing source attribution techniques are designed to operate in a closed-set scenario, i.e. they can only be used to discriminate between known image generators.
By contrast, new image generation techniques are rapidly emerging.  To contend with this, there is a great need for open set source attribution techniques that can identify when synthetic images have originated from new, unseen generators.
 To address this problem, we propose a new metric learning-based approach. 
 Our technique works by learning transferrable embeddings capable of discriminating between generators, even when they are not seen during training.
An image is first assigned to a candidate generator, then is accepted or rejected based on its distance in the embedding space from known generators' learned reference points.
 Importantly, we identify that initializing our source attribution embedding network by pretraining it on image camera identification can improve our embeddings' transferability.
 Through a series of experiments, we demonstrate our approach's ability to attribute the source  of synthetic images in open-set scenarios.

\end{abstract}

\section{Introduction}
\label{sec:intro}

As deep learning techniques have evolved rapidly in  recent years, AI-based image synthesis algorithms have become increasingly successful and ubiquitous.
Researchers have used techniques such as variational autoencoders~\cite{kingma2013auto, rombach2022high}, GAN based generators~\cite{goodfellow2020generative}, diffusion models~\cite{ho2020denoising, ho2022cascaded, rombach2022high}, and other techniques~\cite{gharbi2017deep, zhang2017beyond, Liu_2022_CVPR, mildenhall2020nerf} to generate images that look like they were captured by a real camera.
While some argue that these techniques enable artists be more creative,
synthetic images can also be employed for malicious purposes, including online misinformation and disinformation campaigns.

In order to address the potential threats posed by novel AI-generated multimedia content, researchers have developed a wide variety of highly-performing algorithms aimed at detecting synthetic images. Wang et al. \cite{wang2020cnn} analyzed the periodic signal present in the frequency domain of synthetic images, which is caused by up-sampling in the neural network generator. Other methods have also been proposed for detecting synthetic images using learned forensic traces \cite{Rossler_2019_ICCV}.

However, with the widespread use of synthetic images, it has become increasingly important to trace their sources.
That is because the images' origins
can be used to trace and reveal the nature of a particular misinformation or disinformation campaign. For example, it is important to know if a synthetic image circulated online came from a known source or from a new, unknown origin
To achieve this, researchers have developed several approaches to attribute the source of synthetic images. Frank et al. \cite{frank2020leveraging} proposed using DCT-CNN to discriminate generators from DCT-transformed images. Bui et al. \cite{bui2022repmix} proposed a feature mix-up method to train a synthetic image source attribution model. Albright et al. \cite{albright2019source} proposed an inverse method to identify if an image belongs to the target generator by training proxy models for each of them.


While these approaches have great reported performance,
they are only designed to operate in a closed-set classification scenario
and cannot be easily adapted to work on new sources.
This is problematic because new generators and generation techniques are rapidly emerging.
Therefore, closed-set synthetic source attribution algorithms are not suitable to be used in the real world. And, to the best of our knowledge, comparatively less work had been proposed to solve this urgent and novel problem.
Girish et al. \cite{girish2021towards} proposed an iterative clustering algorithm to cluster a large amount of images from unseen generators into groups. However, this work focuses on grouping a large set of synthetic images into many self-similar clusters (typically many more clusters than there are potential sources). It does not provide an identification criteria to predict whether an image came from a known generator or an unknown one.

In this paper, we propose a new algorithm based on metric learning to perform open set synthetic image source attribution.
Our method involves learning transferable embeddings capable of differentiating between synthetic image generators, including those never encountered during training. We achieve this by identifying reference points in the embedding space associated with each known generator. To attribute a synthetic image to its source, we first determine the nearest candidate class reference to the query image's embedding. We then use the distance between the query image's embedding and the identified class reference to either accept or reject the candidate source. If the distance is below a threshold, we accept the candidate source, otherwise, we identify the image to be from an unknown source generator.
We summarize our contributions as follows:
\begin{itemize}
\itemsep0em
\item We develop a new algorithm that can perform open-set synthetic image attribution and outperforms existing approaches.
\item We develop a new metric-learning based embedding to measure the similarity between synthetic images' source generators.
\item We propose a new reject criteria to determine if a query image is from a new generator.
\item We demonstrate that pretraining the embedding network to perform camera model classification can improve model transferability on images from unseen generators.
\end{itemize}


\begin {comment}
As deep learning techniques evolve rapidly in the recent years, AI-synthesis techniques and contents become more successful and popular throughout the internet.
Researchers have developed variational autoencoder~\cite{kingma2013auto}, GAN based generator~\cite{goodfellow2020generative}, and diffusion model~\cite{ho2020denoising} to generate realistic images that human cannot tell real or fake.
These techniques are widely used and applied in different areas, including image enhancement~\cite{gharbi2017deep} to enhance image quality, image denoising~\cite{zhang2017beyond} to suppress noisy information in the image, super-resolution~\cite{ho2022cascaded} to synthesize detail contents in re-sampled images, image rendering~\cite{Liu_2022_CVPR} to render images in a more efficient and detailed way, Neural Radiance Fields~\cite{mildenhall2020nerf} to render photorealistic novel views, prompt-generative contents~\cite{rombach2022high}, and etc.

Although these techniques improve the quality of synthetic content and facilitate multiple research areas, they also increase the potential threat of false information generation. Deepfake~\cite{} techniques, for example, enable people to synthesize videos with face swapped with another person, which enable false video synthesis and AI-puppeteering.
To combat the threats of noval AI-based generative contents, the vision community proposed multiple algorithms to detect synthetic contents. Many research focus on detecting content inconsistency such like symmetric inconsistency~\cite{}, motion inconsistency in video~\cite{}, and other features~\cite{eyeblink}. However, these features could be diminished as the synthetic techniques get improved.

\begin{figure}[h]
\includegraphics[width=\columnwidth]{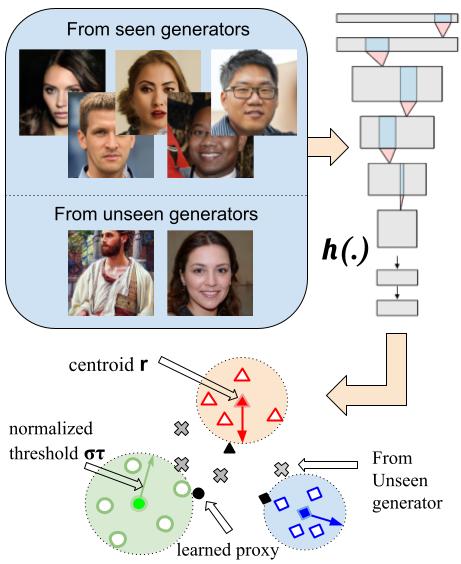}
\caption{Proposed method's pipeline. We proposed a metric-learning based approach for open set synthetic image source attribution. Where images from the seen generators by the algorithm will be accurately classified to be their original generator, and images from unseen generators will be reject by the normalized threshold.}
\end{figure}

Meanwhile, researchers from image forensics community have also proposed multiple approaches. These algorithms are often developed to detect the falsified images, inspired by both traditional signal processing techniques and the recent deep-learning-based algorithms. Traditional image forensic algorithms detect falsified content or traces left by the falsified content in images by extracting hand-craft high-frequency features~\cite{popescu2005exposing, kirchner2008fast, fridrich2012rich, stamm2010forensic, delp2009scanner}.
As deep learning techniques emerge, researchers combine the signal processing analysis with deep learning algorithms and propose multiple solutions. \cite{xu2016structural, tuama2016camera, bayar2018constrained} proposed algorithms using convolutional neural network to learn the features from high-frequency content of images extracted by one or more high-pass filters, which is approved to be efficient on detecting image manipulations. \cite{wang2020cnn} analyze the periodic signal exists in synthetic images, which is caused by up-sampling in the neural network generator. Other methods~\cite{Rossler_2019_ICCV} are also proposed for detecting synthetic image using learned forensic traces. These algorithms shows very strong performance on detecting synthetic images.

Since AI-generative images become widely spread and existing research demonstrate different features from different image synthesis algorithms, algorithms to trace the source of synthetic images become possible and  needed to ensure the authenticity of an image. Researchers develop several approach to attribute the source of synthetic images.\cite{frank2020leveraging} proposed DCT-CNN to discriminate generators from DCT transformed images.\cite{bui2022repmix}  propose a feature mix-up method to train a synthetic image source attribution model.

However, these approaches are trained in a closed-set classification scenario. As new synthetic image techniques emerge rapidly, classification on known generators become unrealistic in real world.
For open-set scenario on synthetic image.
\cite{albright2019source} proposed an inverse method to identify if an image belongs to the target generator, by training proxy models for each new generator, which could be resource-consuming.
\cite{girish2021towards} proposed an iterative clustering algorithms to cluster large amount of images from unseen generators into groups. However, it doesn't perform single image attribution on open-set scenario.
Although there are other open-set image source attribution algorithms have been developed~\cite{mayer2019forensic, huh2018fighting, cozzolino2019noiseprint} and can transfer to image manipulation detection tasks, they are not trained on AI-synthetic images, which have different forensic traces.
To the best of our knowledge, no work has been done specifically for open-set synthetic image source attribution.

In this paper, we proposed a new approach to trace the source of synthetic images in an open-set scenario.
This system includes a neural network trained with metric learning algorithms. The synthetic image's embedding are extracted by the model, then the system compute its distance from each generator's reference point in the embedding space. The reference point with smallest distance is assigned the the synthetic image. After that, the distance value is normalized and compared with a threshold to decide accept or reject by the assign generator class. We note that by pretraining the neural network on images' source camera model classification, the performance of the system on the images from unseen generators can be improved significantly.
Through extensive experiments, We demonstrate our proposed system's capability on open-set synthetic image attribution, and the transferrability introduced by pretraining on camera model datasets.

\end{comment}

\section{Related Work}
\label{sec:related_work}
\subheader{Image Synthesis Algorithms} Researchers proposed multiple image generation techniques, including Variational Auto Encoder~\cite{kingma2013auto}, GAN~\cite{goodfellow2020generative} and the recent diffusion-based models~\cite{ho2020denoising, rombach2022high}. These image synthesis techiques have been widely used to generate different image contents, including prompt-guided image synthesis~\cite{rombach2022high, li2022blip}, and super-resolution~\cite{ho2022cascaded}.

\subheader{Forensic Algorithms} To combat falsified content in images, researchers in the forensic community developed signal processing based methods for forgery detection. These algorithms are often based on human-designed features to detect the inconsistencies in forensic traces in the frequency domain~\cite{popescu2005exposing, kirchner2008fast, fridrich2012rich, stamm2010forensic, delp2009scanner, kirchner2008fast, stamm2008blind, fridrich2008jpeg}.
Additionally, researchers also developed multiple deep-learning based forensic algorithms to detect both traditional image editing~\cite{bayar2016deep, bayar2018constrained}, and AI-generated contents~\cite{Rossler_2019_ICCV}. In particular, \cite{xu2016structural, bayar2018constrained} developed CNNs with high-pass filters to extract generic forensic features from images. This had been proven to be effective at detecting image forgery.

\subheader{Synthetic Image Detection} Researchers have created multiple algorithms to detect traces left by a generator or generator's architecture in synthetic images~\cite{verdoliva2020media}.
Previous research~\cite{wang2020cnn} showed that images from different generators architectures, including deepfake ones, often contain distinctive high-frequency information.
Additionally, other techniques like inversion ~\cite{albright2019source} is also developed for synthetic image detection. This technique works by inverting the generator to obtain the set of features that were used to generate the synthesized image. The authors showed that these features can be used to attribute specific generators.

\subheader{Synthetic Image Source Attribution} Researchers have developed many algorithms for synthetic image attribution~\cite{albright2019source, zhang2020attribution, bui2022repmix}. DCT-CNN~\cite{frank2020leveraging} utilized the high-frequency information by first converting the input image using discrete cosine transform, then used a shallow CNN to perform closed-set source attribution. \cite{bui2022repmix} proposed a new feature mix-up method, and trained the model using a well-balanced training data to perform source attribution in a closed-set scenario. However, currently research on open-set image source attribution remains limited. Sharath el.~\cite{girish2021towards} proposed a method to cluster large amount of images from unseen generators into groups. This is different from source attribution on a single query image,
because it does not have an identification mechanism to predict if an image
came from a known or unknown generator.
To the best of our knowledge, there is no existing algorithm can perform open-set synthetic image source attribution.

\subheader{Other Open-set Image Source Attribution} Forensic researchers developed many open-set image camera model attribution algorithms~\cite{mayer2019forensic, junior2019depth, huh2018fighting, cozzolino2019noiseprint, bayar2018towards}. These algorithms often leveraged the Siamese network architecture, in which a pair of images from the same camera model is labeled similar, while a pair of images from different camera models is labeled different. These algorithms have been proven to be efficient and transferable to multiple different types of image forgery and manipulation.
\section{Proposed Method}
\label{sec:proposed}

In this section, we introduce our proposed approach for open-set synthetic image source attribution.
Assume the investigator has a set of known image generators $\mathbb{G}$ and a set of synthetic images $\mathbb{T}$ created by $\mathbb{G}$. $\mathbb{T}$ is partitioned into disjoint subsets ${\mathbb{T}_i}$, each belongs to a generator architecture $g_i \in \mathbb{G}$. For a good open-set source attribution approach, the algorithm should correctly identify the source of an synthetic image as $g_i$ if it is from a known generator $g_i \in \mathbb{G}$, or attribute the source to be unknown if the origin is an unknown generator $g_u \notin \mathbb{G}$. 
Overall, an open-set algorithm requires a source identification rule $S:\mathbb{X} \rightarrow \mathbb{G}$, which assigns a candidate generator within the known set $\mathbb{G}$
, and a rejection rule $R:\mathbb{X}\times \mathbb{G}\rightarrow \{0,1\}$
where $0$ indicates that the class identified by $S$ should be rejected and the image comes from an unknown generator, and $1$ indicates that the candidate class should be accepted.

While existing closed-set approaches are good at producing $S$, they are not designed to produce $R$.
To address this problem,
we propose a source identification rule $S$ defined by an embedding function $h(\cdot)$, a distance metric $d(\cdot, \cdot)$, and a set of class references ${r_k}$ on the embedding space. For a synthetic image $x$, the source identification rule $S$ is made by measuring the distance between its embedding $h(x)$ and the reference points $r_k$ for each class in the embedding space using $d(\cdot, \cdot)$. The generator associated to the closest reference point to $h(x)$ is identified as the candidate source $g \in \mathbb{G}$.

It is critical that $h$ is generic and can capture forensic traces that can both
accurately discriminate images among known sources and identify images from unknown sources.
To achieve this, we propose a novel training procedure in which we initialize $h$ by pretraining
the network to perform camera model classification using the Camera Model 
Database~\cite{mayer2018learning,mayer2019forensic,mayer2020exposing}.
Previous studies have shown that this process enables the neural network to learn generic embeddings that can be transferred to other forensic tasks~\cite{mayer2018learning}. Our experiments demonstrate that this procedure significantly improves the model's open-set performance.

After making the closed-set source identification decision with the rule $S$, we determine whether to accept or reject the source generator candidate using rule $R$. This is achieved by first normalizing the distance between the query image's embedding $h(x)$ and the reference point $r_i$ based on the chosen generator $g_i$. Then, if the normalized distance exceeds a threshold $\tau$, we reject the source decision
and consider it to be from an unknown generator. Otherwise, we accept and identify it as being from source $g_i$.

\subsection{Transferable Embedding Initialization}
It is crucial that the embedding function $h$ both accurately discriminate between images from known sources and identify those from unknown sources.
%
A common way to learn an embedding is to train a CNN to discriminate between generators in the known set. However, this does not necessarily learn an embedding that transfer well to unknown generators.
To address this issue, we propose novel training procedure, in which we first initialize $h$ by pretraining on the task of camera model identification with large number of classes. By doing so, we enable $h$ to be more sensitive to small differences in forensic traces, and make $h$'s embedding more transferable to images of unknown sources. Our experiments showed that this procedure significantly improves the open-set performance.
\vspace{-6pt}
\begin{align}
\mathcal{L}_{init} = -\sum_k{y_k log(\hat{y}_k)}
\end{align}
After pretraining, we
removed the last classification layer and used the feature maps before the final layer as the initialization for $h$. Following this procedure, we fine-tuned $h$ to learn a more precise embedding that could effectively discriminate between different image generators in the following section.

\begin{figure}
\begin{minipage}[t]{\textwidth}
\includegraphics[width=\textwidth]{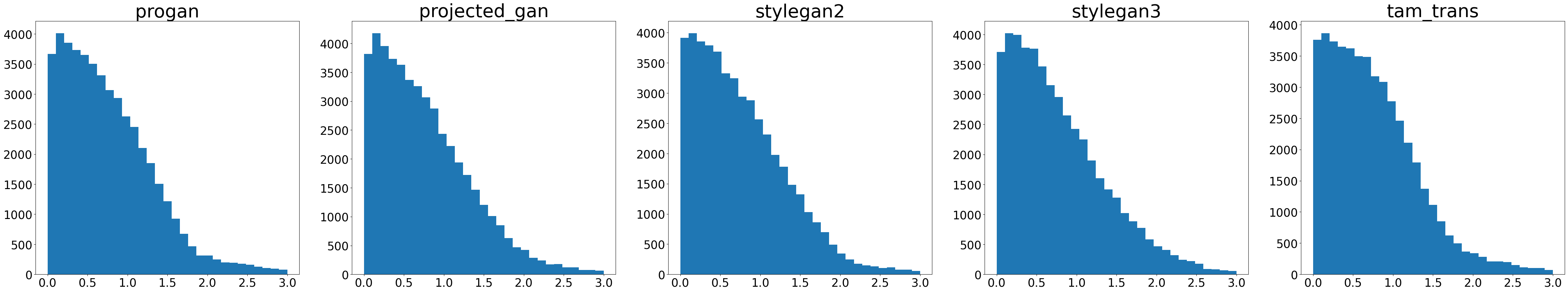}
\caption{Distribution of training set $\mathbb{T}$'s embeddings' normalized distance from reference point. We choose Xception to be the $h$. This figure shows the normalized distance is a good metric for rule $R$ in proposed approach.}
\label{fig:hist}
\vspace{-5pt}
\end{minipage}
\end{figure}

\subsection{Embedding Learning}
Metric learning algorithms enable a model to adapt to unseen types of data during the training process.
These techniques learn an embedding and associated metric such that data points of the same class are closed together and data points of different classes are far apart.
By utilizing this characteristics, we can identify images from unknown generators by testing if their embeddings lie far away from known generators' embedding clusters.
To accomplish this, we utilize ProxyNCA++~\cite{teh2020proxynca} to further train our initialized embedding function $h$. ProxyNCA++ has been widely used to learn embeddings for open set tasks,
and has been shown to have higher accuracy and faster convergence compared to other non-proxy based methods. During training, it uses randomly initialized proxies for each class, pull the embeddings closer to their assigned proxy and push them away from other classes' proxies.
The ProxyNCA++ loss can be written as:
\vspace{-5pt}
\begin{align}
&P_i = \frac{exp(-d(h(x_i), p(y_i)))}{\sum_{p(a)\in p(\mathcal{A})}exp(-d(h(x_i), p(a)))} \\
&\mathcal{L}_{proxyNCA++} = -log(P_i)
\end{align}
where $y_i$ is the generator architecture label of $x_i$, $p(y_i)$ is the corresponding proxy embedding of label $y_i$. The $x$ is the image, $h$ is the feature extractor. Set $\mathcal{A}$ contains all generator classes in the training set. For the distance metric $d(\cdot, \cdot)$, we choose $L_2$ distance to be the metric. 


\subsection{Finding Class References}
After completing the metric learning process and obtaining $h$, we compute the centroid embedding $r_i$ for each class $i$ in the learned embedding space, using all images from generator $g_i$ in the training data $T_i \in \mathbb{T}$.
We compute the centroid of each class' embeddings over the training set as its reference point:
\vspace{-5pt}
\begin{align}
r_i &= \frac{\sum_{x_i\in\mathbb{T}_i}{h(x_i)}}{|\mathbb{T}_i|}
\end{align}
To perform inference, we build $S$ by choosing $x$'s nearest reference point $r_i$ to assign a potential generator class $i$:
\vspace{-5pt}
\begin{align}
S(x) =\operatorname*{argmin}_{k}d(h(x), r_k)
\end{align}
where $d(h(x), r_k)$ is the $L_2$-distance between $x$'s embedding with the reference point $r_k$ of class $k$. After we assign an class $i$, we use a reject criteria to test if $x$ belongs to the class $i$ or is from unseen generator.

\subsection{Reject Criteria}
For open set source attribution, 
we establish a rejection criterion $R$ for all seen classes. We use the normalized distance between $h(x)$ and $r_i$ to determine whether to accept or reject 
the class decision $i$.
This is because each class' embedding group may have different variance, and normalization of the distance enables a better measurement of the relative similarity for each class. The normalized distance $s(x, r_i)$ between $h(x)$ and $r_i$ is defined as:
\vspace{-5pt}
\begin{align}
s(x, r_i) & = \frac{d(h(x), r_i)}{\sigma_i}
\end{align}
where $\sigma$ is computed inspired by sample standard deviation of Gaussian distribution:
\vspace{-5pt}
\begin{align}
\sigma_i&=\sqrt{E(d(h(x_i), r_i)^2)} =\sqrt{E(||h(x_i) - r_i||_2^2)}
   		=\sqrt{\frac{\sum_{x_i\in \mathbb{T}_i}{||h(x_i) - r_i||_2^2}}{|\mathbb{T}_i| - 1}}
\end{align}
This normalization enable $R$ to adapt to variance in the embedding space for each class. Fig~\ref{fig:hist} show san example of the distribution of normalized distances over each $\mathbb{T}_i$ in an embedding space,
using Xception~\cite{chollet2017xception} as the embedding network $h$.
This figure illustrates that normalization can map the distance between $x \in g_i$ and different $r_i$ into a similar distribution.

After that, we define the rejection criterion by comparing the normalized distance $s(x, r_i)$ between $h(x)$ and $r_i$ with a threshold $\tau$. The threshold $\tau$ is applied to all seen classes of the generator. If $s(x, r_i) < \tau_i$, we accept $x_i$ and classify it as coming from a seen generator $g_i$. Otherwise, we reject $x$ and identify it as coming from a new unknown image generator $g_u$:
\vspace{-5pt}
\begin{equation}
 R(x) =
 \begin{cases}
  g_i \in \mathbb{G}, & \text{if}\ s(x,r_i) < \tau \\
  g_u \notin \mathbb{G}, & \text{if}\ s(x,r_i) \geqslant \tau
 \end{cases}
\end{equation}

\section{Experimental Results}
\label{sec:experiments}

This section demonstrates our approach for open-set synthetic image attribution through multiple experiments. We first describe the dataset we create to train and benchmark our proposed algorithm, then we introduce the hyper-parameters we use to train the embedding function $h$. We also propose and discuss the metrics we use for evaluating algorithms' performance. Then, we display our proposed approach's performance on open-set synthetic image attribution.
Lastly, we compare our best model with  state-of-the-art publicly available synthetic image attribution approaches.

The results show that our approach can successfully attribute the source generator's architecture of an image in an open-set scenario. Additionaly, we find that other existing closed-set method cannot be used to analyze images from unseen models, and other existing open-set forensic algorithms do not perform well on synthetic image attribution. Furthermore, we show that pre-training our models on image camera model classification improves there generalizability in attributing synthetic images

\subsection{Dataset}
\begin{table}[t]
     \footnotesize
     \begin{tabular}{lc}
          Generator               & Dataset for Generator Training                       \\
          \midrule
          ProGAN & celebA, lsun-churchoutdoor, lsun-bicycle, lsun-bird, lsun-bedroom, lsun-car  \\
          ProjectedGAN            & ffhq, lsun-bedroom, lsun-churchoutdoor, cityscape    \\
          Tam-Transformer         & ffhq, imagenet-fish, imagenet-shark, imagenet-dog    \\
          StyleGAN2               & ffhq, metfaces, afhqv2, afhqdog                      \\
          StyleGAN3               & ffhq, metfaces, afhqv2                               \\
          \hdashline
          StyleGAN                & ffhq, celebahq, lsun-bedroom, lsun-car               \\
          Stable Diffusion        & N/A                                                 \\
          \bottomrule
     \end{tabular}
     \caption{Generator architectures, and corresponding datasets that are used to train the generator for image synthesizing}
     \label{tab:dataset}
     \vspace{-5pt}
\end{table}

We created our own dataset to train the proposed method.
We used seven publicly available image generators and synthesize multiple images from each. The seven generators are ProGAN~\cite{karras2017progressive}, Projected-GAN~\cite{Sauer2021NEURIPS}, StyleGAN~\cite{karras2019style}, StyleGAN2~\cite{karras2020analyzing}, StyleGAN3~\cite{karras2021alias}, Taming Transformer~\cite{esser2021taming}, and Stable Diffusion~\cite{rombach2022high}. The dataset we used for training the generator are listed in Table.\ref{tab:dataset}. For stable diffusion, we directly used the publicly available  model.

To create our dataset, we generated 10,000 synthetic images for each combination of generator and dataset. Then, for each combination, we picked 8,000 images for training, 1,000 images for validation, and 1,000 images for testing. We hold out images from StyleGAN and Stable Diffusion during training, and used them as the ``unseen" group of generator architecture for evaluating model's performance on the open-set attrbution task.
Since synthetic images are often undergo JPEG compression in the real world,
we randomly JPEG-compressed images using different quality factors of 75, 80, 85, 90, 95, along with 
no compression with equal probability. During training and evaluation, when the image size mismatches with the model's input size, we apply random cropping to 256 by 256 pixels. We intentionally avoid resampling to preserve the forensic features and information of the synthetic images.

\subsection{Training Parameters}
We trained our model using two training protocols to demonstrate the benefits of pre-training on the camera identification task. The first involved training the model from initial weights that were obtained through pre-training, and the second involved training the model from scratch. During all training stages, we apply under-sampling over all seen generators to balance the number of images for each class.

\subheader{Train From Pre-training Initialization}
We chose 70 camera models from 
the Camera Model Identification Database used in~\cite{mayer2018learning,mayer2019forensic,mayer2020exposing} for classification. 
We used AdamW~\cite{loshchilov2017decoupled} optimizer, with an initial learning rate of 0.001, decayed with a scale of 0.65 every 3 epochs. After training, we discarded the classification layer and trained the rest of the model with ProxyNCA++ loss for 30 epochs. The initial learning rate for this stage is 0.0001, which decayed with a scale of 0.6 every 2 epochs.

\subheader{Train From Scratch}
We trained our embedding models from scratch for 30 epochs, still using AdamW~\cite{loshchilov2017decoupled} optimization with an initial learning rate of 7e-4, decay for 0.6 for every 2 epochs. We chose the model with a highest validation accuracy on ``seen'' generator.

\subsection{Evaluation Metrics}
To evaluate different approaches for open-set synthetic image attribution, we used the average F-1 scores ($aF_1$) on seen generator architectures and the correct reject rate (CRR) on unseen architectures.
The $aF_1$ estimates how well the generators perform on seen generator architectures under a specific threshold, balancing the recall and precision for detecting each generator's images.
The $aF_1$ score is defined as the F-1 score averaged over 5 seen $g_s\in\mathbb{G}$:
\begin{equation}
      aF_1 = \sum_{j=1}^N F_{1j} / N,\;\;\;\;\; where\;\;F_{1j} = \frac{2TP_j}{2TP_j + FP_j + FN_j}
\end{equation}
where $TP_j$ is the number of samples from generator architecture $j$ being predicted from generator $j$. $FP_j$ is the number of samples not from generator architecture $j$, including both seen and unseen architectures, to be predicted from generator $j$. $FN_j$ is the number of samples from generator $j$ but are not classifier from $j$. $N$ is the number of seen generators.

The correct reject rate (CRR) measures the ability of the model to reject an image to be from an unseen generator architectures ${g_u \notin \mathbb{G}}$. It is defined as:
\vspace{-5pt}
\begin{align}
     CRR = \frac{|\;\{\min_{i}s(x, r_i) > \tau, \;\;\forall x \in \{g_u \notin \mathbb{G}\}\}\;|}{|\;\{\forall x \in   \{g_u \notin \{\mathbb{G}\}\}\;|}
\end{align}
which is the probability of images from unseen generator architecture $g_u$ (StyleGAN, Stable Diffusion) being rejected. The higher the CRR, the better the discrimination ability of the embedding function $h$.

\subsection{Choosing Optimal Embedding Function}
\begin{table}[t]
	\footnotesize
	\centering
	\setlength{\tabcolsep}{3pt}
     \begin{tabular}{lccccc}
          Embedding Arch. & Xception & ResNet50 & CamID-CNN & Stega-CNN & MISLNet \\
		  \midrule          
          Train From Scratch & 0.827 & 0.671 & 0.519 & 0.787 & 0.761 \\
          With Pre-Training & \textbf{0.868} & 0.714 & 0.574 & 0.808 & \textbf{0.868}\\
          \bottomrule
     \end{tabular}
     \caption{AUC of $aF_1-CRR$ response curve for different models. Pre-training on camera model classification significantly improved the models' generalizability and performance on open-set identification scenario.}
     \label{tab:selection}
\vspace{-5pt}
\end{table}
We conduct a series of experiments to choose the best embedding function $h$.
For the embedding architecture of $h$, we evaluated five different CNNs and compared their performance to determine the best candidate. The five CNNs are Xception~\cite{chollet2017xception}, ResNet50~\cite{he2016deep}, CamID CNN~\cite{tuama2016camera}, Stega-CNN~\cite{zhan2017image} and MISLNet~\cite{bayar2018constrained}. Among them, Xception and ResNet50 were originally designed to perform object recognition tasks, both of them have also widely used by the forensic community on image forensic tasks~\cite{chen2021locally, chen2021robust, gragnaniello2021gan, huh2018fighting, wang2020cnn, bui2022repmix}. Stega-CNN was designed for steganalysis tasks. CamID~CNN and MISLNet were originally designed for camera model identification. We choose these architectures because previous research has shown that they performed well in learning low-level forensic features from images. 

\subheader{Benefit Of Pre-Training On Camera Model Identification}
We conducted an experiment to verify the importance of pre-training our embedding architecture before applying metric learning. We did this by evaluating the AUC of the $aF_1-CRR$ curve for all five embedding architectures with and without camera model identification pre-training. 

From the result displayed in Table \ref{tab:selection}, we see that the pre-training on camera model improved all five model's AUC of $aF_1-CRR$ curve. Among them, Xception and MISLNet achieved the best performance with an AUC of 0.868.
This result demonstrates that pre-training on the camera model classification provides useful prior information to learn synthetic image embeddings and improves the model's transferability in an open-set scenario.


\subheader{Model Selection}
We compared the performance of each embedding architecture with pretraining. From Table~\ref{tab:selection}, we can see that both MISLNet and Xception achieve the highest performance. In the following experiments, we select MISLNet as the embedding function $h$ for further evaluation. We choose MISLNet because it is a very light-weight architecture and can potentially have less over-fitting during training.

\subsection{Open Set Attribution Performance}
\begin{table*}[h]
     \footnotesize
     \setlength{\tabcolsep}{1pt}
     \begin{tabular}{ccccccc|ccc}
          Method    & ProGAN & Proj.-GAN & StyleGAN2 & StyleGAN3 & Taming Trans. & $aF_1$ & StyleGAN & Stable Diffusion & CRR \\
          \toprule
          RepMix
          & 0.669 & 0.827 & 0.762 & 0.839 & 0.860 & 0.791 & 0 & 0 & 0 \\ 

          DCT-CNN
          & 0.673 & 0.929 & 0.687 & 0.609 & 0.851 & 0.750 & 0 & 0 & 0 \\ 

          ResNet-50
          & 0.572 & 0.995 & 0.995 & 0.797 & 0.976 & 0.867 & 0 & 0 & 0 \\ 

          \hdashline

          Proposed
          & 0.744 & 0.974 & 0.875 & 0.969 & 0.940 & \textbf{0.900} & 0.484 & 0.806 & \textbf{0.645} \\ 

          \hdashline

          FSM
          & 0.000 & 0.032 & 0.000 & 0.385 & 0.585 & 0.200 & 0.910 & 0.363 & 0.637 \\ 

          EXIF-Net
          & 0.374 & 0.245 & 0.124 & 0.187 & 0.163 & 0.219 & 0.525 & 0.741 & 0.633 \\ 

          \bottomrule
     \end{tabular}
     \caption{Comparison with other existing approaches on open-set synthetic image attribution. Competing algorithms include: closed-set synthetic image attribution approaches, and open-set image source identification approaches. Results shows that our proposed outperformed both types of algorithms.}
     \label{tab:openset}
\vspace{-5pt}
\end{table*}
We evaluate our proposed approach's ability to perform open set source attribution using our final selection (MISLNet) for the embedding architecture.
We compared our performance to several existing closed set synthetic image source attribution approaches, namely DCT-CNN~\cite{frank2020leveraging} and RepMix~\cite{bui2022repmix}.
Additionally, we trained ResNet-50 to perform synthetic image source attribution, as this network has been widely used in other publications to perform synthetic image detection~\cite{huh2018fighting, wang2020cnn, bui2022repmix}.
We note that to the best of our knowledge, there are no directly comparable open-set source synthetic image source attribution approaches.
We trained these CNNs as classifiers with training set images from 5 known generators.

Furthermore, we compared our performance to several open-set approaches to measure the similarity between forensic traces, namely FSM~\cite{mayer2019forensic} and EXIFNet~\cite{huh2018fighting}.
While not directly trained to perform synthetic image source ID, these networks are designed to determine the similarity of forensic traces between two image patches.
We used the publicly available implementations of these algorithms to perform synthetic image source attribution by applying the same open-set identification mechanism with our approach. To compute the reference point $r_i$ for each class $i$, we computed the average embedding output by the Siamese arm's feature extractor. For $d(\cdot, \cdot)$ we directly used the similarity metric on top of the two siamese arms.
The evaluation result is shown in Table~\ref{tab:openset}. From the results we see that our proposed approach can successfully identify images from known generators and reject images from unknown generators. Our approach also outperforms existing algorithms.

\subheader{Comparison With Closed Set Approaches}
Table~\ref{tab:openset} shows the resulting comparison of our approach and other existing closed-set approaches (RepMix, DCT-CNN, and ResNet-50). We achieved higher $aF_1$ on the set of seen generator architectures than competing closed-set approaches.  Furthermore, our approach obtained an average CRR of 0.645 while these methods can only get a CRR of 0.  This shows that in open set scenarios, we are able to outperform these techniques in terms of both $aF_1$ and CRR.  We note that we can have even higher $aF_1$ score at the cost of lowered CRR. 

\subheader{Comparison With Open Set Forensic Algorithms}
From Table~\ref{tab:openset}, we see that our proposed method obtained a higher CRR than existing open-set approaches.  Furthermore, we achieved a substantially higher $aF_1$ on the set of known sources than these networks. This demonstrates that while existing open-set approaches are able to reject images to be from an unknown generator architecture, they have very little ability to correctly identify known sources. Hence, existing open-set forensic algorithms cannot adapt to the synthetic images attribution task. However, our proposed method can both reliably identify if an image comes from a generator architecture that was seen during training, and reject an image coming from an unknown architecture.
\section{Conclusion}
\label{sec:conclusion}

In this paper, we proposed a new algorithm to perform open-set synthetic image attribution.
Through extensive experiments, we demonstrate that our system can successfully perform open-set synthetic image attribution, and outperforms existing methods.
In our approach, we use metric-learning to learn an embedding, and compare synthetic images' source generators by distance measuring its embedding's distance from a class reference in the embedding space. 
We also propose a new accept/reject criteria for images from unseen generators in open-set scenario.
Additionally, we demonstrate that  pre-training an embedding network to perform camera model identification helps improve its transferrability to unknown generators when performing  synthetic image attribution. 
Through a set of experiments, we verify the importance of embedding pre-training and show that our proposed approach can succesfully perform open set synthetic image source attribution.



\bibliography{egbib}
\end{document}